\documentclass[%
reprint,
superscriptaddress,
amsmath,amssymb,
aps,longbibliography]{revtex4-1}
\usepackage{graphicx}
\usepackage{dsfont}
\usepackage{dcolumn}
\usepackage{bm}
\usepackage{lipsum}
\usepackage{color}
\usepackage[normalem]{ulem}
\usepackage{simplewick}
\usepackage{qcircuit}
\usepackage{braket}
\usepackage{float}
\usepackage{multirow}
\usepackage{tikz}
\usepackage[utf8]{inputenc}
\usepackage{braket}
\usepackage{subfig}
\usepackage{makecell}

\begin{document}
\title{Quantum equation of motion for computing molecular excitation energies on a noisy quantum processor}
\author{Pauline J. Ollitrault}
\affiliation{IBM Research GmbH, Zurich Research Laboratory, S\"aumerstrasse 4, 8803 R\"uschlikon, Switzerland}
\affiliation{Laboratory of Physical Chemistry,
ETH Z\"urich, 8093 Z\"urich, Switzerland}
\author{Abhinav Kandala}
\affiliation{IBM T.J. Watson Research Center, Yorktown Heights, NY 10598, USA}
\author{Chun-Fu Chen}
\affiliation{IBM T.J. Watson Research Center, Yorktown Heights, NY 10598, USA}
\author{Panagiotis Kl. Barkoutsos}
\affiliation{IBM Research GmbH, Zurich Research Laboratory, S\"aumerstrasse 4, 8803 R\"uschlikon, Switzerland}
\author{Antonio Mezzacapo}
\affiliation{IBM T.J. Watson Research Center, Yorktown Heights, NY 10598, USA}
\author{Marco Pistoia}
\affiliation{IBM T.J. Watson Research Center, Yorktown Heights, NY 10598, USA}
\author{Sarah Sheldon}
\affiliation{IBM T.J. Watson Research Center, Yorktown Heights, NY 10598, USA}
\author{Stefan Woerner}
\affiliation{IBM Research GmbH, Zurich Research Laboratory, S\"aumerstrasse 4, 8803 R\"uschlikon, Switzerland}
\author{Jay M. Gambetta}
\affiliation{IBM T.J. Watson Research Center, Yorktown Heights, NY 10598, USA}
\author{Ivano Tavernelli}
\email{ita@zurich.ibm.com}
\affiliation{IBM Research GmbH, Zurich Research Laboratory, S\"aumerstrasse 4, 8803 R\"uschlikon, Switzerland}
\date{\today}
\begin{abstract}
The computation of molecular excitation energies is essential for predicting photo-induced reactions of chemical and technological interest.
While the classical computing resources needed for this task scale poorly, quantum algorithms emerge as promising alternatives.
In particular, the extension of the variational quantum eigensolver algorithm to the computation of the excitation energies is an attractive option. 
However, there is currently a lack of such algorithms for correlated molecular systems that is amenable to near-term, noisy hardware. 
In this work, we propose an extension of the well-established classical equation of motion approach to a quantum algorithm for the calculation of molecular excitation energies on noisy quantum computers.
In particular, we demonstrate the efficiency of this approach in the calculation of the excitation energies of the LiH molecule on an IBM Quantum computer.
\end{abstract}

\pacs{Valid PACS appear here}
\maketitle
\section{Introduction}
Quantum computing is emerging as a new paradigm to solve a wide class of problems that show unfavorable scaling on conventional classical high performance computers \cite{Glaesemann2010, Dreuw2005}. In particular, solving quantum chemistry and quantum physics problems using classical algorithms is hampered by the exponential growth of the resources (classical processors and memory) as a function of the number of degrees of freedom, $N$, (e.g., number of electrons or molecular basis functions) encoded in the system Hamiltonian.\\
\indent The resources needed to compute the solution of the Schr\"odinger equation for molecular and solid state systems have a factorial scaling in the full Configuartion Interaction (full CI) representation of the ground state wave function \footnote{The correct scaling will be $\begin{pmatrix} N_{b} \\ N_{\rm el}\end{pmatrix}$, where $N_{b}$ is the number of basis functions and $N_{\rm el}$ is the number of electrons.}.
At present, the canonical Coupled Cluster (CC) expansion truncated at the second order in the electronic excitation operator and including an approximated treatment of the triple excitations (CCSD(T), S stands for single, D
for double, and (T) for non-iterative triple)~\cite{Raghavachari1989, Bartlett2007} with a scaling $\mathcal{O}(N^7)$ is often considered to be the ``gold standard'' for quantum chemistry calculations. Energies computed at the canonical CCSD(T) level of theory have an error that lies within the so-called chemical accuracy (errors less than 1 kcal/mol = 0.043 eV) for many systems (i.e., when no strong static correlation or multi-reference character of the ground state is present~\cite{Becke2013, Ziesche1997}).\\
\indent Recently, the Variational Quantum Eigensolver (VQE) algorithm \cite{Peruzzo2014, Yung2014, Mcclean2016, Wang2019} was proposed for the efficient approximation of the electronic structure in near-term quantum computers. This algorithm is based on a parametrization of trial ground state wave functions. The parameters are encoded in single-qubit and two-qubit gate angles and are optimized self-consistently, using a classical processor, until the minimum ground state energy is reached.
The energy corresponding to a given set of parameters is obtained by computing the expectation value of the system Hamiltonian and becomes therefore a function of the gate variables.
The VQE has already been successfully applied to the simulation of the ground state properties of simple molecular systems on quantum hardware \cite{Kandala2017, Kandala2019,Nam2019}, and extended to more complex molecules in quantum simulators \cite{Sokolov2020, Gao2019}.\\
\indent The calculation of molecular excited state properties constitutes an additional challenge for both classical and quantum electronic structure algorithms. In fact, in addition to calculating a well-converged ground state wave function, one needs to devise schemes for the evaluation of the higher energy states, which – in general – are not accessible through the optimization of a trial state.
In classical computing, excited states are typically computed in linear response theory, explicitly (LR) or implicitly (the equation of motion, EOM) starting from a ground state wave function optimized at a given level of theory (e.g., CC, multi-configurartional self-consistent field, configuration interaction, etc. \cite{Glaesemann2010}). 
In particular, CC theory was also extended to the calculation of excited state wave functions and energies using, for example, LR \cite{Monkhorst1977}, EOM \cite{Stanton1993}, state-universal multi-reference \cite{Jeziorski1981} and valence-universal multi-reference approaches \cite{Lyakh2011}.
Alternatively, density functional theory (DFT) in its time-dependent formulation (namely, time-dependent density functional theory, TDDFT) or Green’s function based techniques (like the Bethe-Salpeter equation, BSE) can be used to evaluate excited state properties in the LR regime.
Since the EOM approach is based on a direct treatment of the ground state wave function, it is an interesting choice for the implementation of excited state properties within the VQE approach \cite{Ganzhorn2019}.\\
\indent Other VQE based quantum algorithms for computing electronic transition energies were recently proposed~\cite{Parrish2019, nakanishi2019, ryabinkin2018, higgott2019, Mcclean2017, stair2019, santagati2018, jones2019, tilly2020, Peruzzo2014}.
A useful overview of these algorithms can be found in \cite{Parrish2019}. 
Among these algorithms the Quantum Subspace Expansion (QSE) \cite{Mcclean2017} was employed to compute the excited states of the $\rm{H_2}$ molecule using two qubits \cite{Colless2018}. 
The advantage of the QSE algorithm compared to the more recent ones relies on the fact that it can be implemented as a straightforward extension of the VQE algorithm as it does not require any modification of the quantum circuit but rather only the measurement of additional matrix elements for the ground state wave function. \\
\indent In this work, we show how to adapt the EOM method to a quantum algorithm to efficiently calculate molecular excitation energies. 
We show how this leads to similar working equations as for QSE and we highlight the differences between both approaches. 
In a second part, we discuss the numerical simulations performed to test our algorithm on several small molecules, namely ${\rm H_2}$, ${\rm LiH}$ and ${\rm H_2O}$. 
Most importantly, we achieve excited states experiments for the ${\rm LiH}$ molecule on the 20 qubit processor IBM Q Poughkeepsie. To do so we vary the noise level in the quantum hardware and adapt an error mitigation technique to this purpose. With this work we demonstrate the robustness of the EOM approach in a noisy quantum computation.\\
\begin{figure*}[t]
  \includegraphics[width = 0.9\textwidth]{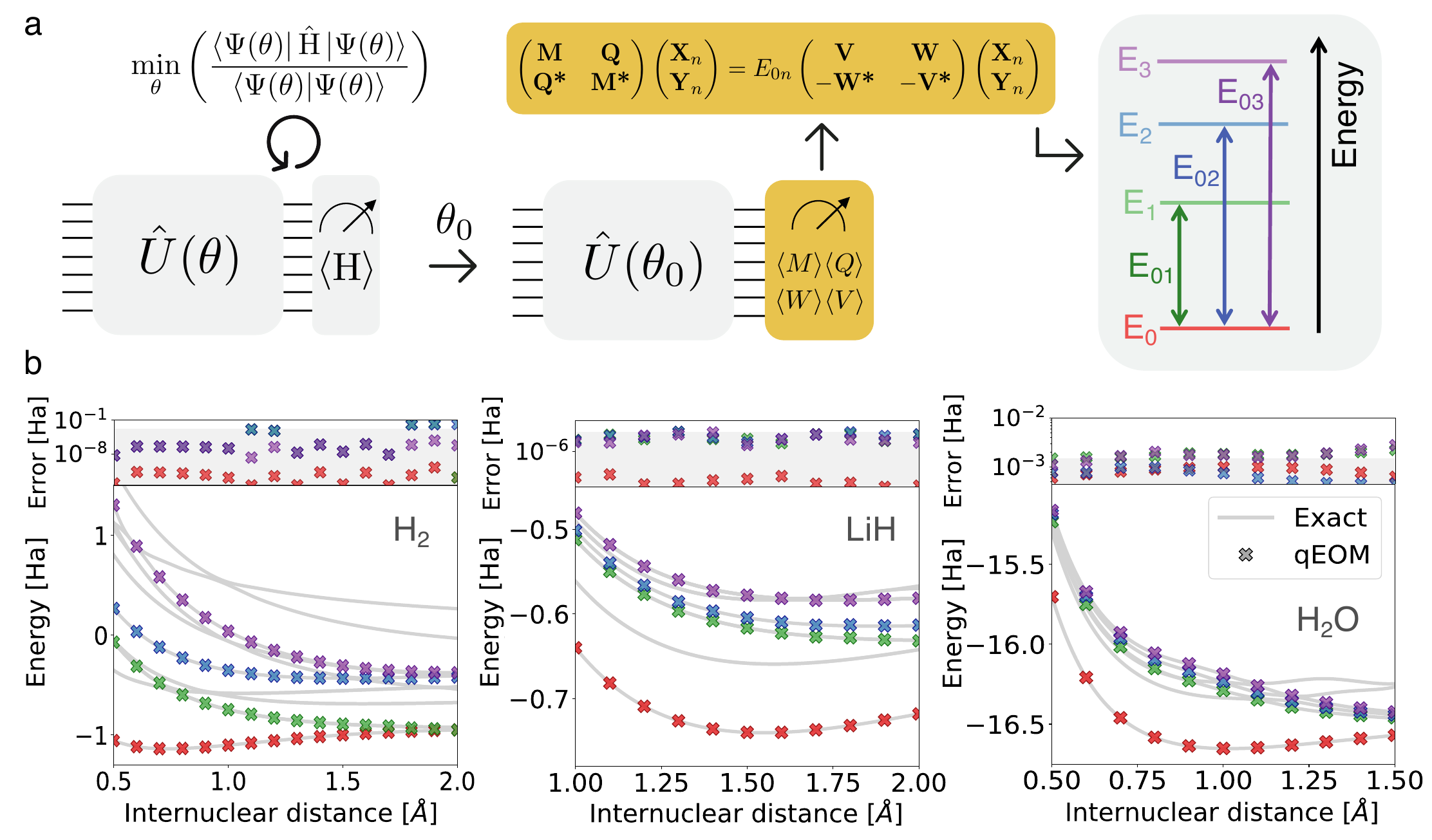}
  \caption{\textbf{a.} Graphical representation of the qEOM algorithm.
  \textbf{b.} Lower panels: Dissociation profile of the $\rm{H_2}$, the $\rm{LiH}$ and the $\rm{H_2O}$ molecules.
  The gray lines represent the exact eigenenergies of the Hamiltonian obtained from its diagonalizaton. The coloured crosses depict the ground state (red),
  the first (green), the second (blue) and the third (purple) excited states obtained with the qEOM. 
  Upper panels: Corresponding energy errors along the dissociation profile. The gray shaded area corresponds to the energy range within chemical accuracy. }
  \label{fig:DISS}
\end{figure*}
\section{Theoretical foundation}
\label{sec:theory}
The EOM approach, first derived by Rowe~\cite{Rowe1968}, was extensively reviewed~\cite{Mcweeny1992, Mccurdy1977} and implemented in a series of electronic structure packages.  
Within this approach, excited states $\ket{n}$ are generated by applying an excitation operator
of the general form
$ \hat{O}_n^{\dagger}=\ket{n}\bra{0} $
to the ground state $\ket{0}$ of the system, where $\ket{n}$ is the shorthand notation for the $n$-th excited state of the electronic structure Hamiltonian.
Similarly, a de-excitation operator can be written as $\hat{O}_n=\ket{0}\bra{n}$.\\
\indent Taking the commutator of the Hamiltonian and the excitation operator leads to an expression for the excitation energies, $ E_{0n} = E_n - E_0$,
\begin{equation}
    [\hat{H},\hat{O}_n^{\dagger}]\ket{0} = \hat{H}\hat{O}_n^{\dagger}\ket{0} - \hat{O}_n^{\dagger}\hat{H}\ket{0} = E_{0n}\hat{O}_n^{\dagger}\ket{0}.
    \label{eigenoperator1}
\end{equation}
Operating from the left hand side with  $\hat{O}_n^{\dagger}\ket{0}$ we then obtain
\begin{equation}
E_{0n} = \frac{\bra{0} [\hat{O}_n, [\hat{\text{H}}, \hat{O} ^{\dagger}_n]] \ket{0}}{\bra{0} [\hat{O}_n, \hat{O}^{\dagger}_n] \ket{0}} = \frac{\bra{0} [\hat{O}_n, \hat{\text{H}}, \hat{O} ^{\dagger}_n] \ket{0}}{\bra{0} [\hat{O}_n, \hat{O}^{\dagger}_n] \ket{0}}.
\label{eq:variational}
\end{equation}
The second equality arises because for the exact ground state, $\bra{0}[\hat{O}_n, [\hat{\text{H}}, \hat{O} ^{\dagger}_n]]\ket{0} = \bra{0}[[\hat{O}_n, \hat{\text{H}}],\hat{O} ^{\dagger}_n]\ket{0}$, and therefore we can symmetrize the numerator introducing the double commutator $[\hat{\text{A}}, \hat{\text{B}}, \hat{\text{C}}] = \frac{1}{2}\{ [[\hat{\text{A}}, \hat{\text{B}}], \hat{\text{C}}] + [\hat{\text{A}}, [\hat{\text{B}}, \hat{\text{C}}]] \}$.~\footnote{This operation has also the important effect of guaranteeing  real valued energy differences $E_{0n}$. In fact, while $\bra{0} [\hat{O}_n, [\hat{\text{H}}, \hat{O} ^{\dagger}_n]] \ket{0}$ may not be Hermitian, the double commutator of the right hand side is Hermitian.} 
Note that this expression differs from the one derived for QSE~\cite{Mcclean2017} 
in two points: (i) due to the commutator form, the operators in the numerator and denominator are Hermitian 
allowing for a systematic reduction of the number of terms to evaluate thanks to the use of the Pauli commutation relations~\cite{hemmatiyan2018},
and (ii) the solution of the EOM equations leads directly to the excitation energies rather than the absolute energies, making the approach size-intensive (contrary to QSE, which is not size-intensive).\\
The EOM approach aims at finding approximate solutions to Eq.~\eqref{eq:variational} by expressing $\hat{O}_n^{\dagger}$ as a linear combination of basis excitation operators with variable expansion coefficients. The excitation energies are then obtained through the minimization of Eq.~\eqref{eq:variational} in the coefficient space. 
The simplest basis is composed of the Fermionic orbital creation and annihilation operators $\hat{a}^{\dagger}$ and $\hat{a}$, where $\hat{a}_m^{\dagger} \hat{a}_i$ represents
 the excitation of a single electron from an occupied orbital $i$ to a virtual orbital $m$, and $\hat{a}_m^{\dagger} \hat{a}_n^{\dagger} \hat{a}_i \hat{a}_j$ the double excitation of a pair of electrons from the occupied orbitals $i, j$ to the virtual orbitals $m, n$. 
 Calling $\alpha$ the degree of excitation, we can express $\hat{O}_{n}^{\dagger}$ as
\begin{equation} \label{eq:O_expansion}
    \hat{O}_{n}^{\dagger} = \sum_{\alpha} \sum_{\mu_{\alpha}} [X_{\mu_{\alpha}}^{(\alpha)} (n) \hat{\text{E}}_{\mu_{\alpha}}^{(\alpha)} - Y_{\mu_{\alpha}}^{(\alpha)}(n) (\hat{\text{E}}_{\mu_{\alpha}}^{(\alpha)})^{\dagger}] \, ,
\end{equation}
where $\mu_{\alpha}$ is a collective index for all one-electron orbitals involved in the excitation.
This expression is general and explicitly treats the possible de-excitation of all states involved. This is of particular importance in the context of quantum computing where the prepared ground state is, in general, a many-determinant wave function. 
Note also that the QSE approach introduced in~\cite{Mcclean2017} neglects the de-excitation operators (as in the Tamm-Dancoff approximation~\cite{Fetter2012, Ring2004}), which limits its application to systems for which this approximation is valid (e.g., far from conical intersections, see for instance~\cite{shu2017}).
\indent In this work, we will restrict our excitation operator basis to single ($\alpha = 1$) and double ($\alpha = 2$) excitations such that $\hat{\text{E}}_{\mu_1}^{(1)} = \hat{a}_m^{\dagger} \hat{a}_i$, $\hat{\text{E}}_{\mu_2}^{(2)} = \hat{a}_m^{\dagger} \hat{a}_n^{\dagger} \hat{a}_i \hat{a}_j $, $(\hat{\text{E}}_{\mu_1}^{(1)})^{\dagger} = \hat{a}_i^{\dagger} \hat{a}_m$ and $(\hat{\text{E}}_{\mu_2}^{(2)})^{\dagger} = \hat{a}_i^{\dagger} \hat{a}_j^{\dagger} \hat{a}_m \hat{a}_n$. 
By inserting the expansion of Eq.~\eqref{eq:O_expansion} into Eq.~\eqref{eq:variational} we obtain a parametric equation for the excitation energies. 
Applying the variational principle $\delta(E_{0n})=0$ in the parameter space spanned by the coefficients $X_{\mu_{\alpha}}^{(\alpha)}$ and $Y_{\mu_{\alpha}}^{(\alpha)}$ we obtain the following secular equation
\begin{equation} 
\begin{pmatrix}
    \text{\textbf{M}} & \text{\textbf{Q}}\\ 
    \text{\textbf{Q*}} & \text{\textbf{M*}}
\end{pmatrix}
\begin{pmatrix}
    \text{\textbf{X}}_n\\ 
    \text{\textbf{Y}}_n
\end{pmatrix}
= E_{0n}
\begin{pmatrix}
    \text{\textbf{V}} & \text{\textbf{W}}\\ 
    -\text{\textbf{W*}} & -\text{\textbf{V*}}
\end{pmatrix}
\begin{pmatrix}
    \text{\textbf{X}}_n\\ 
    \text{\textbf{Y}}_n
\end{pmatrix}
,
\label{eq:XnYn}
\end{equation}
where
\begin{align}
&M_{\mu_{\alpha}\nu_{\beta}} = \bra{0} [(\hat{\text{E}}_{\mu_{\alpha}}^{(\alpha)})^{\dagger},\hat{\text{H}}, \hat{\text{E}}_{\nu_{\beta}}^{(\beta)}]\ket{0}, \nonumber\\
& Q_{\mu_{\alpha}\nu_{\beta}} = -\bra{0} [(\hat{\text{E}}_{\mu_{\alpha}}^{(\alpha)})^{\dagger}, \hat{\text{H}}, (\hat{\text{E}}_{\nu_{\beta}}^{(\beta)})^{\dagger}]\ket{0},\nonumber\\
& V_{\mu_{\alpha}\nu_{\beta}} = \bra{0} [(\hat{\text{E}}_{\mu_{\alpha}}^{(\alpha)})^{\dagger}, \hat{\text{E}}_{\nu_{\beta}}^{(\beta)}]\ket{0},\nonumber\\
&W_{\mu_{\alpha}\nu_{\beta}} = -\bra{0} [(\hat{\text{E}}_{\mu_\alpha}^{(\alpha)})^{\dagger}, (\hat{\text{E}}_{\nu_{\beta}}^{(\beta)})^{\dagger}]\ket{0}.\nonumber
\end{align}
Note that the rank of all these matrices equals the number of possible single and double excitations included in the active space that defines the operators in Eq.~\eqref{eq:O_expansion}. 
Due to the small rank of the matrices involved in the solution of EOM equations, the eigenvalues of Eq.~\eqref{eq:XnYn} can be evaluated classically. 
However, quantum advantage can be achieved through the efficient measurement of each single matrix element of the EOM generalized eigenvalue problem. 
Indeed, while classically the scaling of this operation depends on the wave-function Ansatz, the measurement of the expectation values in a quantum computer scales with the number of terms in the Hamiltonian as $\mathcal{O}(N^4)$.
The steps of this quantum algorithm are the following.
First, the Jordan-Wigner transformation can be used to map the commutators $[(\hat{\text{E}}_{\mu_{\alpha}}^{(\alpha)})^{\dagger},\hat{\text{H}}, \hat{\text{E}}_{\nu_{\beta}}^{(\beta)}]$, $[(\hat{\text{E}}_{\mu_{\alpha}}^{(\alpha)})^{\dagger}, \hat{\text{H}}, (\hat{\text{E}}_{\nu_{\beta}}^{(\beta)})^{\dagger}]$, $[(\hat{\text{E}}_{\mu_{\alpha}}^{(\alpha)})^{\dagger}, \hat{\text{E}}_{\nu_{\beta}}^{(\beta)}]$ and $[(\hat{\text{E}}_{\mu_\alpha}^{(\alpha)})^{\dagger}, (\hat{\text{E}}_{\nu_{\beta}}^{(\beta)})^{\dagger}]$, which are originally expressed in terms of the Fermionic creation and annihilation operators, into the qubits space.
They are then evaluated using the ground state wave function prepared in the quantum hardware from, e.g., a VQE calculation, to compute the matrix elements of \textbf{M}, \textbf{Q}, \textbf{V} and \textbf{W}. 
From these measurements the secular equation (Eq.~\eqref{eq:XnYn}) is constructed. Its $2n$ eigenvalues are then classically solved to obtain the first $n$ excitation (and corresponding de-excitation) energies.\\
Note that the implementation of the EOM approach as a quantum algorithm differs from its classical counterpart and therefore we do not expect the quantum-EOM algorithm to reproduce the same results. 
In fact, the way the EOM matrix elements are evaluated, the different nature of wavefunction Ansatz (i.e., CCSD vs. UCCSD) and its implementation as a quantum circuit, as well as the noise of the quantum processors will always introduce differences in the numerical outcomes compared to the classical solution.
For these reasons, in the following we will use the acronym qEOM to designate the quantum implementation of the EOM algorithm (and not the formal theoretical development, which is the same as in the classical case).
A graphical representation of the qEOM algorithm is given in Fig.~\ref{fig:DISS}a.
It could be further improved by using a classical iterative diagonalization method such as the ones employed in the efficient variants of the classical EOM algorithms, e.g. the Stanton-Bartlett approach~\cite{Stanton1993}
(with scaling $\mathcal{O}(N^6)$). Also in this case, the advantage of the qEOM algorithm resides in the efficient evaluation of the required expectation values.
\section{Simulations of the qEOM algorithm}
\label{sec:Simulations}
To validate the performance and the accuracy of the qEOM approach, statevector-type simulations (where the exact unitary matrix representation of the circuit is applied on the state vector, no sampling or hardware noise is included) are performed.
The algorithm is tested on three molecules, namely ${\rm H_2}$, ${\rm LiH}$ and ${\rm H_2O}$. 
We prepare the Hamiltonians by computing the one- and two-electron integrals using a Hartree-Fock/STO-3G calculation performed on Gaussian09~\cite{g09d01}.
To improve the efficiency of the ground state calculation, the resulting Hamiltonian is mapped into the particle-hole framework~\cite{Barkoutsos2018}.
In the case of ${\rm LiH}$ and ${\rm H_2O}$ effective core potentials (ECPs)~\cite{StuttgartEcp} are used to replace $1s$ core electrons. 
In this way, we are able to reduce the number of qubits for LiH and H$_2$O resulting in 4, 10 and 12 qubits for H$_2$, LiH and H$_2$O, respectively.\\ 
\indent In Fig.~\ref{fig:DISS}b we show the dissociation profile of ${\rm H_2}$, ${\rm LiH}$ and ${\rm H_2O}$ obtained with qEOM (coloured crosses) and from the exact diagonalization of the Hamiltonian (gray lines). 
The qEOM operators are measured on a ground state obtained with VQE. The wave function is approximated with the UCCSD Ansatz where the indices for the excitations run over all occupied and all virtual Hartree-Fock orbitals for the given basis set. The variational parameters are optimized with the L-BFGS-B algorithm as implemented in SciPy \cite{2020SciPy}. 
The excited states shown in Fig.~\ref{fig:DISS} are the 3 lowest lying excited states found by solving the qEOM pseudo-eigenvalue equation. 
In this work, we choose to use only particle and spin conserving excitation operators $\hat{\text{E}}_{\mu_{\alpha}}^{(\alpha)}$. For this reason, the qEOM excited states displayed in Fig.~\ref{fig:DISS} are not strictly the 3 lowest lying ones but rather the 3 lowest lying within this specific particle- and spin-number subspace. 
By selecting a specific subset of the excitation operators it is therefore possible to address specific sectors of interest of the entire Hilbert space.\\
\indent In the case of $\rm H_2$ and $\rm LiH$ the accuracy of the ground state is excellent over the entire dissociation curve (see upper panels in Fig.~\ref{fig:DISS}), allowing us to compute accurate excited state energies within chemical accuracy (with errors $< 0.015$ Hartree, shaded area) for all geometries. 
For H$_2$O, the VQE results are less accurate, leading to excited states slightly above chemical accuracy. \\
\indent Assessing the error propagation from the ground state wave function calculation to the excited states energies is not straightforward and we propose a more detailed discussion in the Supplementary Materials.
We found the error in the excitation energies to grow more slowly than the error in the ground state energy when adding an increasing error to the ground state parameters. 
The accuracy of the excitation energy is altered because of the propagation of errors through Eq.~\eqref{eq:XnYn} due to the nature of the \textbf{M}, \textbf{V}, \textbf{Q} and \textbf{W} matrices.
We expect the sampling noise to have a stronger effect on the results in strong correlation regimes. 
Moreover, the propagation of errors from the ground state to the excitation energies was demonstrated to be weaker with qEOM than QSE. 
However, when the ground and first excited states become almost degenerate, i.e., when the lowest energy solution of the EOM generalized eigenvalue problem, Eq.~\eqref{eq:XnYn}, is close to zero, the conditioning of the corresponding matrices deteriorates, affecting the quality of the resulting excitation energies.\\
\indent In preparation of computing molecular excited states on a quantum hardware, we implement the qEOM algorithm in the Qiskit software library~\cite{Qiskit}. 
This enables us to perform realistic noisy simulations and model the performance of the algorithm on the IBM Q Poughkeepsie device. 
Using Qiskit, the molecular orbitals are computed with Hartree-Fock/STO-3G on the PySCF classical code \cite{PYSCF}. In this case the core orbitals are simply frozen (note that the absolute energies are shifted in comparison to using ECPs). The circuit depth is reduced as explained in the next section.
We observe that around the optimal ground state parameter value the error in the excitation energies is about one order of magnitude smaller than in the ground state energy (see Supplementary Materials). \\
In the next section the relative robustness of the experimentally obtained excitation energies measured at varying noise levels is studied. 
\begin{figure}[t]
  \includegraphics[width = \columnwidth]{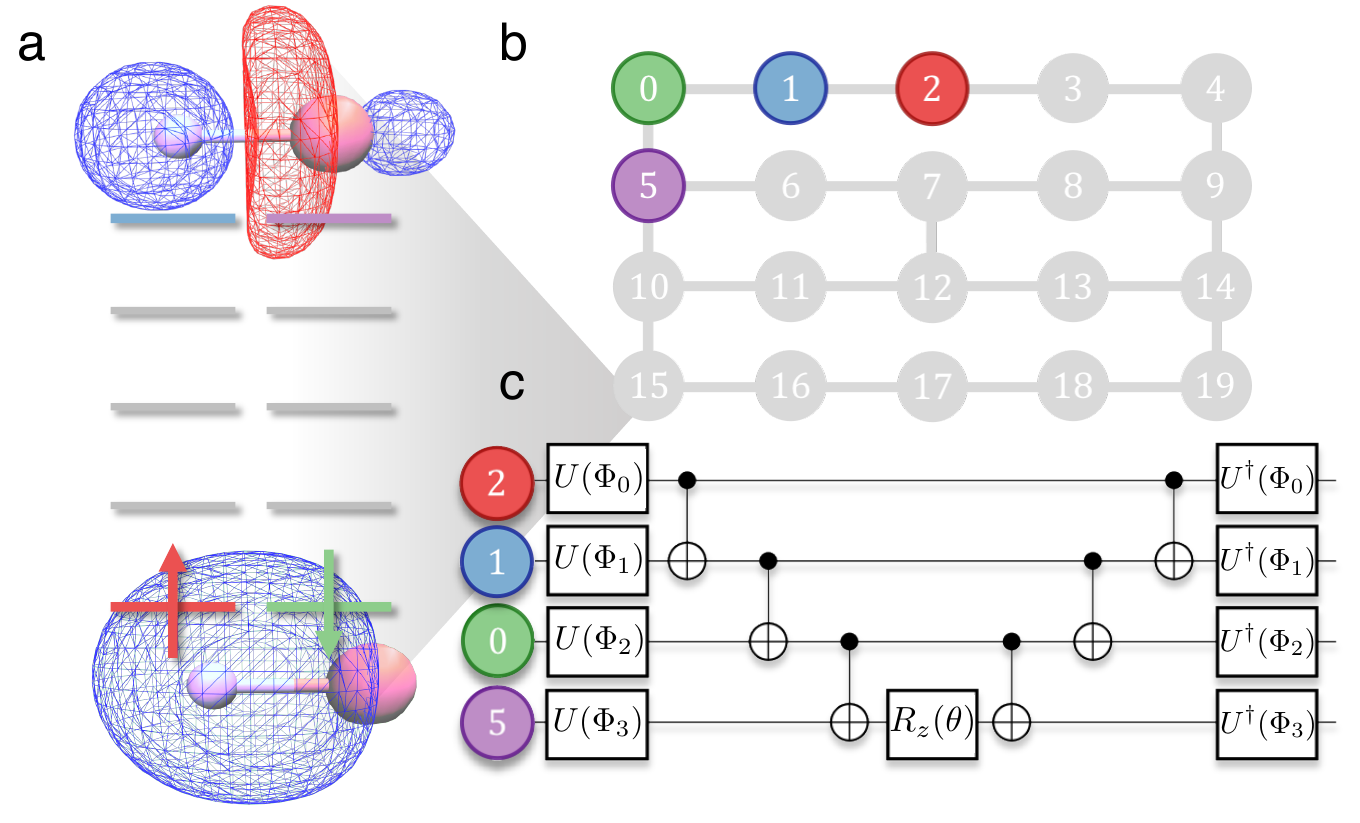}
  \caption{\textbf{a.} Active space and corresponding molecular orbitals of LiH. \textbf{b.} IBM Q Poughkeepsie device layout. The active orbitals are mapped onto the coloured qubits. \textbf{c.} UCC-inspired circuit of 4 qubits. The sets $\Phi_i$ of angles are: $\Phi_0=\{\pi/2,0.0,0.930\}$, $\Phi_1=\{-\pi/2,\pi,-1.207\}$, $\Phi_2=\{-\pi/2,-\pi,1.310\}$,$\Phi_3=\{-\pi/2,0.0,1.877\}$.}
  \label{fig:active_space}
\end{figure}
\section{Hardware calculation of the excitation energies of LiH}
\label{sec:hardware}
\begin{figure*}[t]
  \includegraphics[width = 0.9\textwidth]{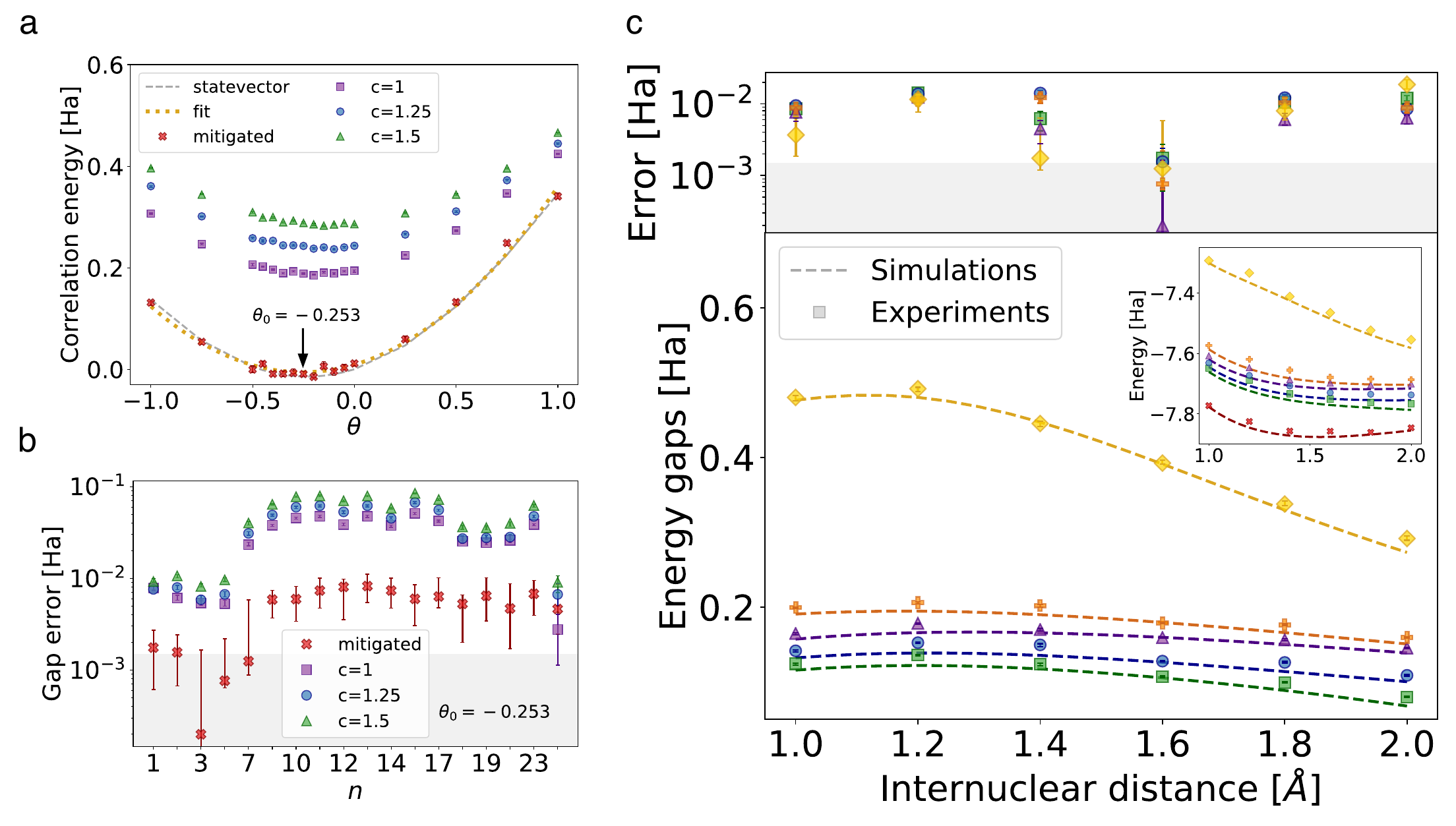}
  \caption{\textbf{a.} Ground state correlation energy versus the variational parameter $\theta$ at equilibrium bond length (1.6\AA). The energies computed with statevector-type simulations are shown together with the experimental results obtain for each of the three stretch factors and after mitigation. The yellow dotted curve displays the fit to the mitigated values. \textbf{b.} Error in the $n$ energy gaps at equilibrium bond length (1.6\AA). The error is computed with respect to the results obtained by statevector-type simulation of the reduced circuit. The results are shown for each of the three stretch factors and after mitigation. The energy of the gaps grows from left to right. \textbf{c.} Lower panel: Dissociation profile of the five lowest-lying electronic transition energies of the LiH molecule. The markers show the mitigated experimental results. The dashed lines are the qEOM results from statevector-type simulations (using the reduced 4-qubit active space). 
  Inset: Dissociation profile of the 6 lowest-lying electronic states of the LiH molecule. The qEOM transition energies are added to the ground state energy (red).
  Upper pannel: Corresponding energy errors along the dissociation profile. The gray shaded area corresponds to the energy range within chemical accuracy. The errorbars are computed using 50 numerical experiments obtained by bootstrapping of the experimental data points, and depict the range between the 1st and 3rd quantile.}
  \label{fig:es_diss}
\end{figure*}
The UCCSD circuit for the optimization of the LiH ground state using a STO-3G basis set comprises over 12000 CNOT gates and 92 variational parameters. 
Given the limitations of state-of-the-art quantum hardware, we reduce the active space from 10 to 4 orbitals.
The reduction of the active space in the quantum computing framework has already been discussed in literature~\cite{Barkoutsos2018,Takeshita2019}.
The orbitals composing the active space in LiH are selected according to their contribution to the CI expansion (see Supplementary Materials).
The resulting active space is shown in Fig.~\ref{fig:active_space} along with the layout of the 20 qubit superconducting processor IBM Q Poughkeepsie used for the experiment.
In the `conventional' quantum UCCSD, double excitations are encoded using 8 entangling blocks with different fixed pre/post rotations~\cite{Romero2018,Barkoutsos2018}. 
Here, we replace the 8 blocks by a single one with variable pre/post rotations (in $U$) where the angles are optimized in simulation to best approximate the exact UCCSD results (see Supplementary Materials). 
Due to this pre-processing procedure we are able to reduce the circuit, for LiH, to 6 CNOTs, 8 fixed single qubit rotations and a single variable qubit rotation $R_z(\theta)$ 
as shown in Fig.~\ref{fig:active_space}.
This modified variational UCC circuit used in the reduced active space can recover at least 56\% of the correlation energy (and up to 87\%, see the details in the Supplementary Materials) over the entire dissociation range considered, which corresponds to an energy error $\leq 7$ mHa. 
Discrepancies are expected to be larger for long internuclear distances ($>2.5$\AA) where strongly correlated effects become more important and a larger active space is therefore required.\\
\begin{table*}
	\begin{tabular*}{6.5in}{l|@{\extracolsep{\fill}}*{6}{c}}
		\hline \hline
		Gate 		           & $c=1$ 	    & $c=1.25$ 	    & $c=1.5$ 	    	\\ \hline
		$\text{Q}_0 \text{ simultaneous}$& $0.0016\pm4.2\times10^{-5}$ 		&	$0.0023\pm8.3\times10^{-5}$ 		&	$0.0028\pm 8.3\times10^{-5}$ 	
		\\ \hline
		$\text{Q}_1 \text{ simultaneous}$  & $0.0007\pm1.7\times10^{-5} $  & $0.0010\pm2.7\times10^{-5}$  & $0.0012\pm 3.1\times10^{-5}$  
		\\ \hline
		$\text{Q}_2 \text{ simultaneous}$  & $0.0011\pm2.3\times10^{-5} $  & $0.0013\pm2.9\times10^{-5}$  & $0.0018\pm 4.2\times10^{-5}$  
		\\ \hline
		$\text{Q}_5 \text{ simultaneous}$  & $0.0008\pm1.7\times10^{-5} $  & $0.0011\pm2.0\times10^{-5}$  & $0.0013 \pm 2.5\times10^{-5}$   
		\\ \hline
		$\text{CNOT}_{2-1}$ &	0.031 $\pm$ 0.001		& 0.030 $\pm$ 0.001		&	0.032 $\pm$ 0.001	 
		\\ \hline
		$\text{CNOT}_{1-0}$ &	0.037 $\pm$ 0.001		& 0.038 $\pm$ 0.001		&	0.042 $\pm$ 0.001
		\\ \hline
		$\text{CNOT}_{0-5}$ &	0.038 $\pm$ 0.001		& 0.043 $\pm$ 0.002		&	0.048 $\pm$ 0.002	 
		\\ \hline \hline 					
	\end{tabular*}
		\caption{\label{table:device_parameter} \textbf{Gate characterization} Single and two qubit gate fidelities for the gates employed in this work, for the various stretch factors, estimated by randomized benchmarking.}
\end{table*}
The experiments presented in this work used 4 superconducting qubits (Q0, Q1 ,Q2 and Q5). See Fig.~\ref{fig:active_space} for the connectivity of the 20 qubit processor IBM Q Poughkeepsie. The qubit frequencies are in the range 4.8-5 GHz, with relaxation and coherence times of T$_1$ and T$_{2,\rm{echo}} \sim 40-110 \mu$s. The single and two-qubit gates are implemented by all-microwave drives. Every trial circuit is composed of 6 CNOT gates, implemented using cross-resonance pulses and single qubit gates. The shortest single qubit gates used for the experiments are of duration 103 ns, and the shortest gate times for CNOT$_{2_1}$, CNOT$_{1_0}$ and CNOT$_{0_5}$ are 1278 ns, 1210 ns and 1448 ns respectively. To improve the quality of the computation, we use the error mitigation scheme previously implemented in~\cite{Kandala2019}. Here, expectation values of interest are re-measured under amplified noise strengths in order to then extrapolate to the zero-noise limit. Under the assumption of time invariant noise, this noise amplification is achieved by stretching in time the single and two qubit gates that constitute the quantum circuit of interest. In this work, we employ stretch factors of $c=1,1.25,1.5$ and use linear extrapolation for obtaining zero-noise estimates. An important consideration for noise amplification by stretching the gates is the introduction of undesired coherent errors, that could result in unphysical extrapolations. In this context, we employ a four-pulse echo sequence for the construction of the $ZX_{90}$  gate that serves as the primitive for realizing a CNOT. Similar sequences have in the past been employed to mitigate the effect of spectator interactions in parity measurements for quantum error correction~\cite{Takita2016}. At each stretch factor, the gates are characterized by randomized benchmarking and the obtained fiedlities are reported in Table 1. The average readout assignment errors for the four qubits were $\epsilon_r \sim 0.05$. As discussed in \cite{Kandala2017,Kandala2019} all measured expectation values were corrected for assignment infidelity using a readout calibration of all the basis states.
\indent At each internuclear distance, the circuit parameter $\theta$ is swept, and energies are measured at three different stretch factors (1.0, 1.25 and 1.5), and a mitigated sweep is obtain using a linear extrapolation of these measurements. To reduce the sensitivity to any fluctuations, the optimal $\theta$ for the ground-state is obtained by fitting a quadratic curve to the mitigated sweep, as shown in Fig.~\ref{fig:es_diss}a at the equilibrium bond length. The error mitigation protocol is then extended to each matrix element of the qEOM and the resulting ``mitigated'' secular equation is then solved classically, leading to the excitation energies $E_{0n}$ with $n=1,\dots,24$. The absolute errors with respect to the statevector-type simulations are displayed on Fig.~\ref{fig:es_diss}b. Error mitigation enables a gain in precision of approximately one order of magnitude in both the ground state and excitation energies. More importantly, we experimentally observe that the unmitigated results are more accurate for the excited states than the ground state (about 1e-2 Ha for the lowest energy states against an error superior to $\sim$1e-1 Ha for the ground state), from the runs at different stretch/noise amplification factors. Finally, we test our algorithm for varying internuclear distance, discussed in Fig.~\ref{fig:es_diss}c (and Supplementary Materials).
\section{Conclusions}
\label{sec:conclusion}
In this work, we introduced a quantum algorithm for the calculation of electronic excited state energies based on the classical equation of motion approach (EOM). 
The method, named qEOM, inherits the benefits of the variational approaches for ground state calculations as well as all merits of the EOM method, while improving upon its classical counterpart by taking advantage of the efficient measurement of expectation values in a quantum computer.
We tested the qEOM algorithm on three small molecules: $\rm{H_2}$, $\rm{LiH}$ and $\rm{H_2O}$, demonstrating that simulations can produce excitation energies within chemical accuracy (errors $\leq1.5$ mH).
We studied the performance of our algorithm and showed that it is particularly well suited for calculations on state-of-the-art quantum device, manifesting robustness against hardware noise. 
Finally, we adapted an error mitigation scheme to the qEOM approach and were able to compute the excitation energies of LiH on the IBM Q Poughkeepsie device. 
The stability of the qEOM algorithm, demonstrated in this work, opens up new avenues in the use of quantum computers for studying photochemical processes. \\
\section*{Acknowledgements}
The authors thank Igor Sokolov, Guglielmo Mazzola, Christa Zoufal, Almudena Carrera Vazquez, Marc Ganzhorn, Daniel Egger, Stefan Fillip, Nikolaj Moll, Ali Javadi-Abhari, Ken X. Wei and Kristan Temme for useful discussions as well as Isaac Lauer, Douglas T. McClure, Srikanth Srinivasan and Neereja Sundaresan for experimental contributions.\\
The authors acknowledge financial support from the Swiss National Science Foundation (SNF) through the grant No. 200021-179312.\\
IBM, the IBM logo, and ibm.com are trademarks of International Business Machines Corp., registered in many jurisdictions worldwide. Other product and service names might be trademarks of IBM or other companies. The current list of IBM trademarks is available at https://www.ibm.com/legal/copytrade.\\
A.M. acknowledges support from the IBM Research Frontiers Institute.\\
\section*{Author contributions}
I.T. and S.W. conceived the idea. I.T. supervised the theoretical part and J.M.G. the experimental part of the work. P.J.O. developed the algorithm. P.J.O., P.Kl.B and A.M. ran the simulations. A.K., P.J.O., C.C, A.M, S.S., J.M.G and I.T. designed the experiments. A.K. and S.S. performed the experiments. C.C., A.M., M.P. and P.J.O. implemented the algorithm in Qiskit Aqua. All authors contributed to the analysis of the results. P.J.O. and I.T. wrote the first draft of the manuscript and all authors contributed to its final version.
%
%

\clearpage
\widetext
\begin{center}
\textbf{\large Supplementary Materials for:
Quantum equation of motion for computing molecular excitation energies on a noisy quantum processor}
\end{center}
\setcounter{equation}{0}
\setcounter{section}{0}
\setcounter{figure}{0}
\setcounter{table}{0}
\setcounter{page}{1}
\renewcommand{\theequation}{S\arabic{equation}}
\renewcommand{\thefigure}{S\arabic{figure}}
\renewcommand{\bibnumfmt}[1]{[#1]}
\renewcommand{\citenumfont}[1]{#1}
\section{Error propagation}
\label{section:appendix_error}
\begin{figure}[b]
\centering
\includegraphics[width = 0.9\textwidth]{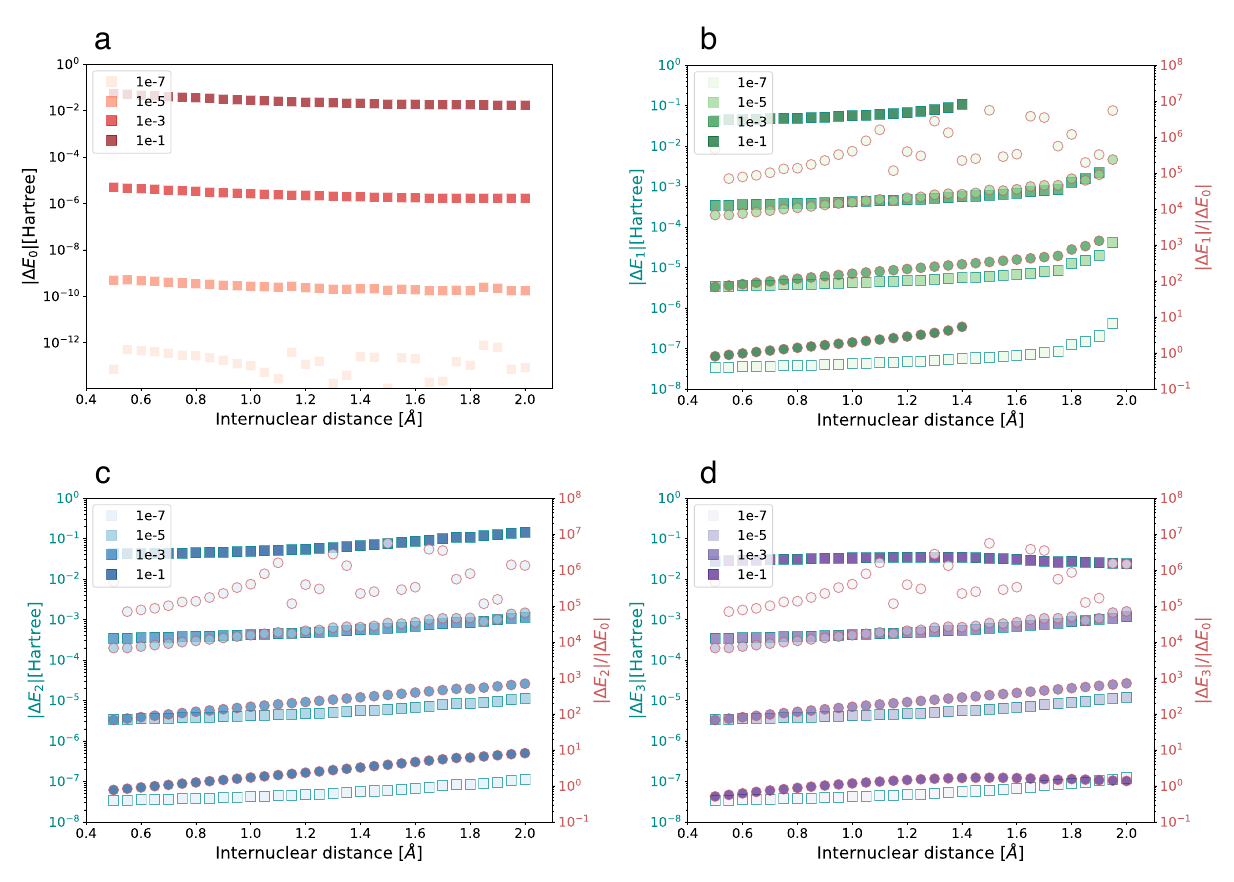}
\caption {Error in \textbf{a} ground state, \textbf{b} first, \textbf{c} second and \textbf{d} third excited state energies when adding an error to the VQE optimized ground state parameters. The legend shows the value of the added error. Squares: absolute difference between the energies computed with and without adding an error to the parameters. Circles: ratio between the absolute errors made on the excited state $i$ and on the ground state.}
\label{fig:ERR_PARAM}
\end{figure}
Assessing the error propagation from the ground state wave function calculation to the excited state energies is not straightforward. 
To shed light on this issue, we compute the ground and excited states of a $\rm{H_2}$ molecule. The Hamiltonian is prepared by computing the one- and two-electron integrals using a Hartree-Fock/STO-3G calculation performed on Gaussian09~\cite{g09d01}. We compute the correlation energy by running a VQE with the UCCSD ansatz in a statevector-type manner (the exact unitary matrix of the circuit is applied on the quantum state vector). For this system, the VQE comprises three variational parameters. 
The excited state energies, $E_i$, are computed by finding the excitation energies (energy gaps) with qEOM and add them to the ground state energy, $E_0$. 
We add an `ad-hoc' error, $\epsilon$, to all parameters defining the ground state wave function and re-compute the ground state as well as the excited state energies as a function of $\epsilon$. 
The absolute value of the corresponding ground and excited state energy variations $|\Delta E_{i}(\epsilon)|=|E_{i}(0)-E_{i}(\epsilon)|$
are reported in Fig.~\ref{fig:ERR_PARAM}\textbf{a}-\textbf{d} for $i \in\{0,1,2,3\}$. Here, $E_{0}$ refers to the ground state energy and $E_{i}$ to the energy of state $i$.
For all excited states calculations, Fig.~\ref{fig:ERR_PARAM}\textbf{b}-\textbf{d} also report the ratio between the errors of the excited state $i$ and the one obtained for the ground state as a function of $\epsilon$.
In all cases, the $|\Delta E_{i}(\epsilon)|$ (with $i\in{0,1,2,3}$) grow monotonically with the increase of $\epsilon$.
Moreover, the error slightly increases with the stretching of the bond. 
Interestingly, we also observe that the larger the value of $\epsilon$ is, the closer the errors in ground and excited states get (circles).
This implies that the qEOM energy gaps are less sensitive to an error in the ground state than the ground state energy itself which, in this case, becomes the main source of error. 
As shown in Fig.~\ref{fig:ERR_PARAM}\textbf{b} it was not possible to obtain energies for the first excited state for $\epsilon=10^{-1}$ and bond lengths larger than 1.4~\AA. 
At these bond lengths the ground and the first excited state energies become almost degenerate. Within these conditions, the system to solve becomes ill-conditioned, and numerical instabilities appear. 
The same argument can also explain the slight deterioration of the first excited state energies for internuclear bond distances larger than $1.75$\AA\ at all levels of noise.
\begin{figure}[t]
\centering
\includegraphics[width = 1\textwidth]{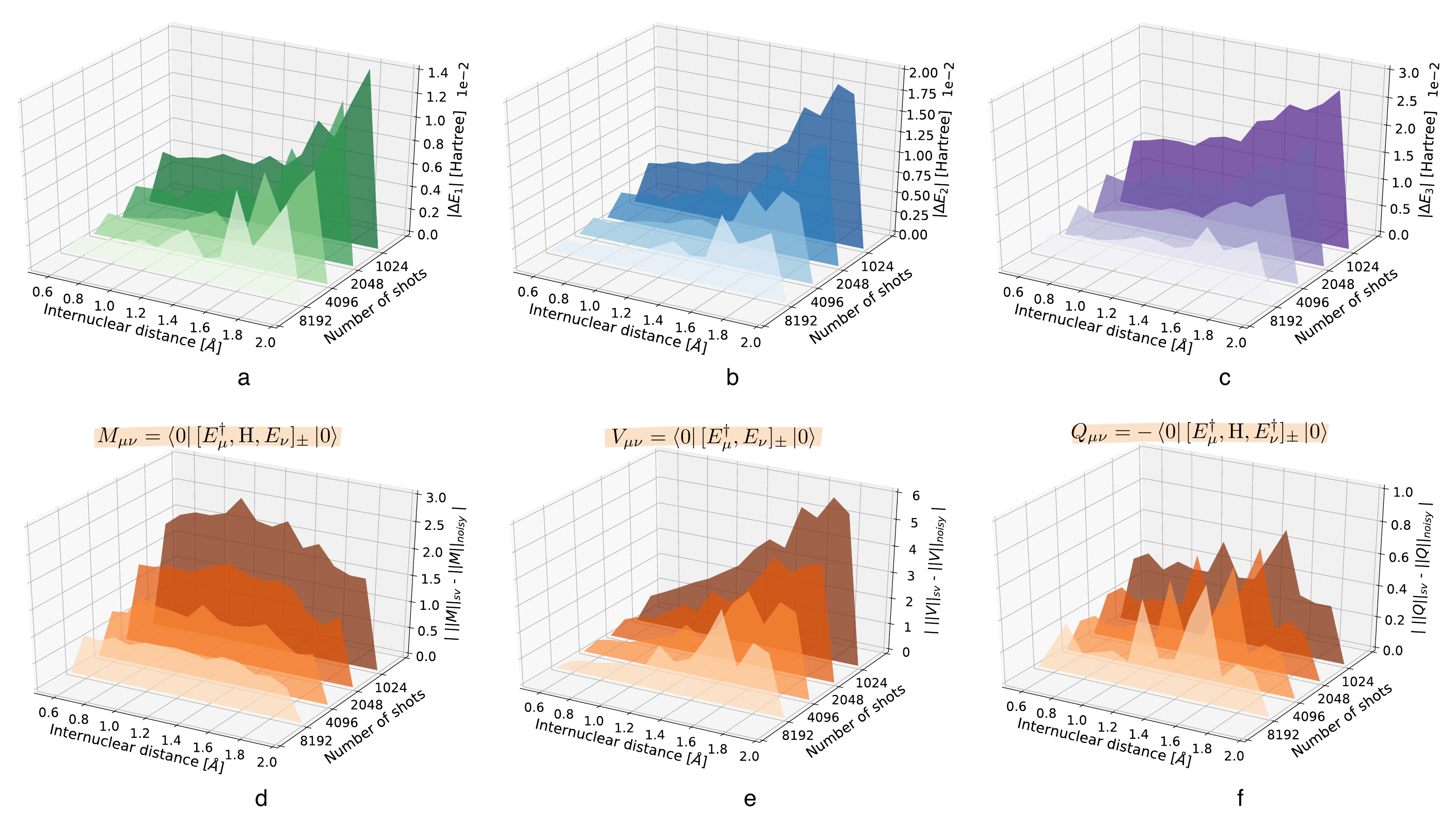}
\caption{Top: absolute error in the \textbf{a} first, \textbf{b} second and \textbf{c} third excited state energies due to shot noise. The error corresponds to the difference between energies computed with and without sampling.
Bottom: absolute error in the norm of the \textbf{d} M, \textbf{e} V and \textbf{f} Q matrices due to finite number of shots. The error corresponds to the difference between the matrix norms computed with and without sampling.}
\label{fig:ERR_NOISE}
\end{figure}
In the following we study the robustness of qEOM to statistical errors. 
The ground state wave function of the $\rm{H_2}$ molecule optimized with a statevector-type VQE (as explained in the previous paragraph) is used to compute the excited state energies with qEOM by introducing this time a statistical error: the expectation value of each matrix element, $M_{\mu_{\alpha}\nu_{\beta}}$, $Q_{\mu_{\alpha}\nu_{\beta}}$, $V_{\mu_{\alpha}\nu_{\beta}}$, and $W_{\mu_{\alpha}\nu_{\beta}}$, is obtained by projecting the state and averaging over a given number of shots, NS $\in \{8192, 4096, 2048, 1024\}$. 
For each different choice of NS we perform 100 computations of excited state energies and compute the error, $|\Delta E|$, with the corresponding values obtained without statistical noise. 
In Fig.~\ref{fig:ERR_NOISE} (top), we plot the average of this error for each bond length for \textbf{a} the first, \textbf{b} the second, and \textbf{c} the third excited state. 
In the second line of Fig.~\ref{fig:ERR_NOISE}, we report the error of the norm of the matrices \textbf{M}, \textbf{V} and \textbf{Q} instead. Note that for H$_2$, \textbf{W} is a null matrix. All absolute errors reported in Fig.~\ref{fig:ERR_NOISE} have been averaged over the aforementioned 100 realizations of the experiment.
In general, we observe an increase of the absolute energy error with the statistical noise, i.e., with the decrease of NS. 
As mentioned above, each element of the \textbf{M}, \textbf{V} and \textbf{Q} matrices corresponds to a weighted sum of PS. 
The energy presents a large variance when the coefficients associated to the PS are large.
In the case of the \textbf{Q} matrix, the PS coefficients are relatively small ($<10$) across the whole range of internuclear distances. The matrix elements $Q_{\mu_{\alpha}\nu_{\beta}}$ are therefore, weakly affected by the sampling error.
The effects on the matrix \textbf{V} are more interesting. 
By definition, the coefficients weighing the PS of the \textbf{V} matrix elements do not depend on the bond length (they do not depend on the Hamiltonian). 
The error in the \textbf{V} matrix elements through the dissociation curve shows that the required number of shots for sampling the distribution increases with the bond length. With internuclear distance, the correlation effects increase and the wave function distribution broadens, requiring more shots to be accurately described. This is translated directly to the accuracy of the \textbf{V} matrix elements. 
Finally, the \textbf{M} matrix elements are weighted by large coefficients ($\sim 50$) at short bond lengths but they decrease as the internuclear distance increases. 
In this case the increase of the PS measurement errors and the decrease of the variance at large bond distances cancel out leading to a roughly constant accuracy for the matrix elements $M_{\mu_{\alpha}\nu_{\beta}}$ over the entire dissociation range. 
We observe that the error in the excited state energies follow the error in the matrix norms (the \textbf{V} matrix, in this case, which is mostly affected). 
We also expect qEOM to be less accurate in strong correlation regimes (e.g. at large bond lengths), when the correct description of the ground state becomes difficult. 
\begin{figure}[t]
\centering
\includegraphics[width = 1\textwidth]{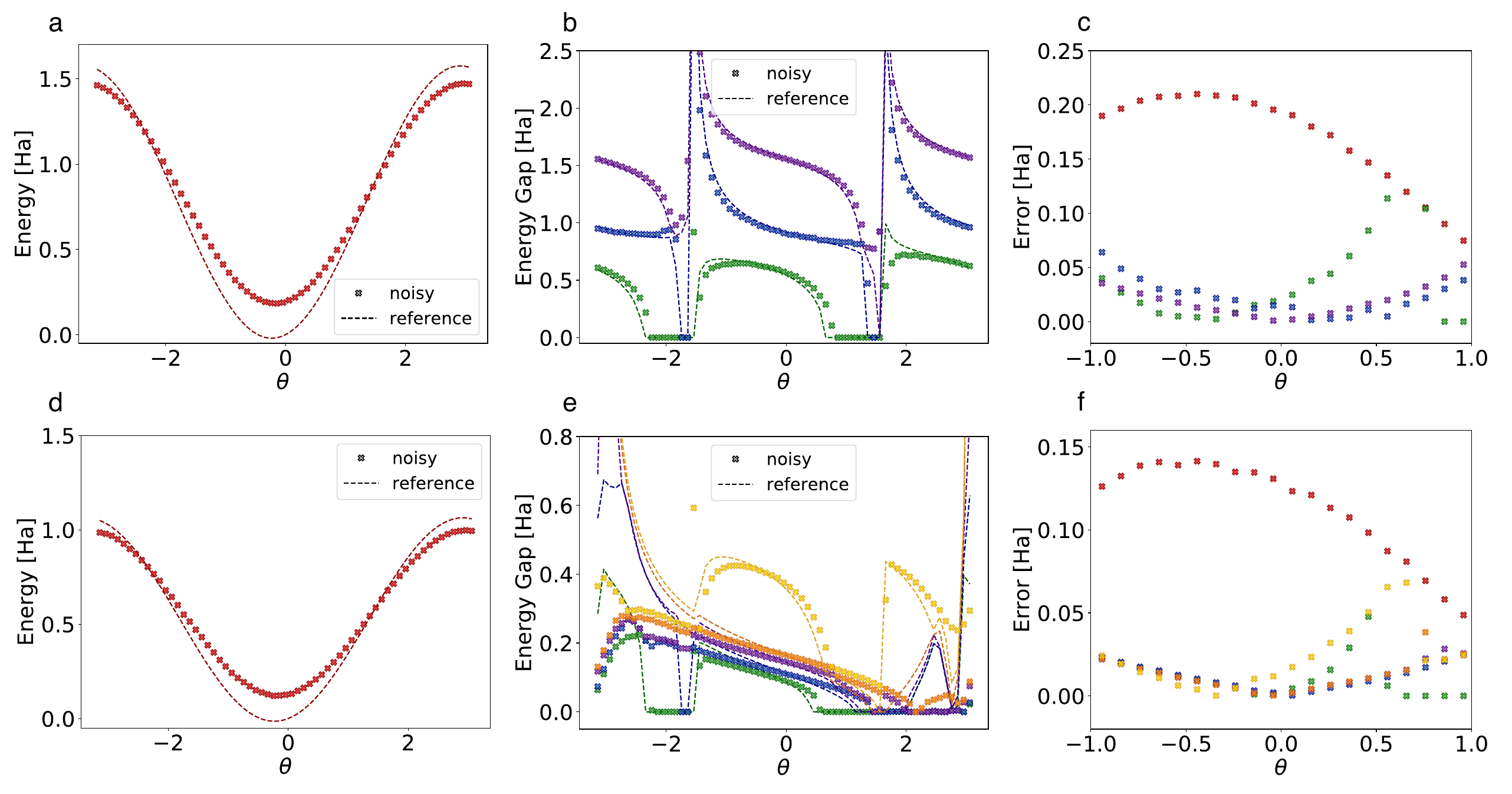}
\caption{Left: Correlation energy of the ground state computed with (markers) and without (dashed lines) noise as a function of the variational parameter $\theta$ for \textbf{a} $\rm{H_2}$ and \textbf{d} $\rm{LiH}$ at equilibrium distance.
Middle: Corresponding excitation energies computed with qEOM with (markers) and without (dashed lines) noise. For $\rm{H_2}$ \textbf{b} the three excitation energies are shown. For $\rm{LiH}$ \textbf{e} the excitation energies corresponding to the five lowest energy states (degeneracies are not taken into account) are displayed.
Right: Absolute error in the noisy simulations with respect to the reference for the ground state and corresponding excitation energies with a zoom in the region $\theta \in [-1;1]$ for \textbf{c} $\rm{H_2}$ and \textbf{f} $\rm{LiH}$}
\label{fig:theta_scan}
\end{figure}
In preparation of computing molecular excited states on a quantum hardware, we implement the qEOM algorithm in the Qiskit software library~\cite{Qiskit}. 
This enables us to perform realistic noisy simulations and model the performance of the algorithm on the IBM Q Poughkeepsie device. 
Using Qiskit, the molecular orbitals of a $\rm{H_2}$ and $\rm{LiH}$ molecules are computed with Hartree-Fock/STO-3G and the PySCF classical code \cite{PYSCF}. The core orbitals are frozen. The circuit depth is reduced as explained in the next section leading to circuits parametrized with a single angle $\theta$ for both molecules. 
The simulations are done using the \textit{qasm} simulator with the Poughkeepsie noise model and 100K shots. 
We switch this parameter within the range $[-\pi;\pi]$ and compute the ground and excited states of $\rm{H_2}$ and $\rm{LiH}$ at equilibrium distance (0.75\AA\ and 1.6\AA\ respectively).  
The effects on the ground state energy as well as the excitation energies (energy gaps directly computed with qEOM) are shown in Fig.~\ref{fig:theta_scan}. 
Around the best $\theta$ value, the noise has the effect to shift up the computed ground state energy. On the other hand the qEOM is robust to the noise, leading to an error in the excitation energies (energy gaps) of about one order of magnitude smaller than for the ground state (see plots \textbf{c} and \textbf{f} of Fig.~\ref{fig:theta_scan}). 
The theta scans (plots \textbf{a} and \textbf{d} of Fig.~\ref{fig:theta_scan}) also help us determine a suitable region to inspect experimentally. In this region the energy change with $\theta$ should be higher than the fluctuations coming from the hardware noise but close enough to the bottom of the well such that the data can be fitted to an harmonic curve. 
Applying a fit to determine the bottom of the curve has the purpose of obtaining a value which is not biased by the fluctuations coming from the hardware (that are in the range of the energy change when approaching the optimal $\theta$ value). This region was selected to be $\theta \in [-1;1]$.
\section{Comparison with QSE}
\label{section:eom_vs_qse}

\begin{figure}[t]
    \subfloat{\includegraphics[width = 0.4\textwidth]{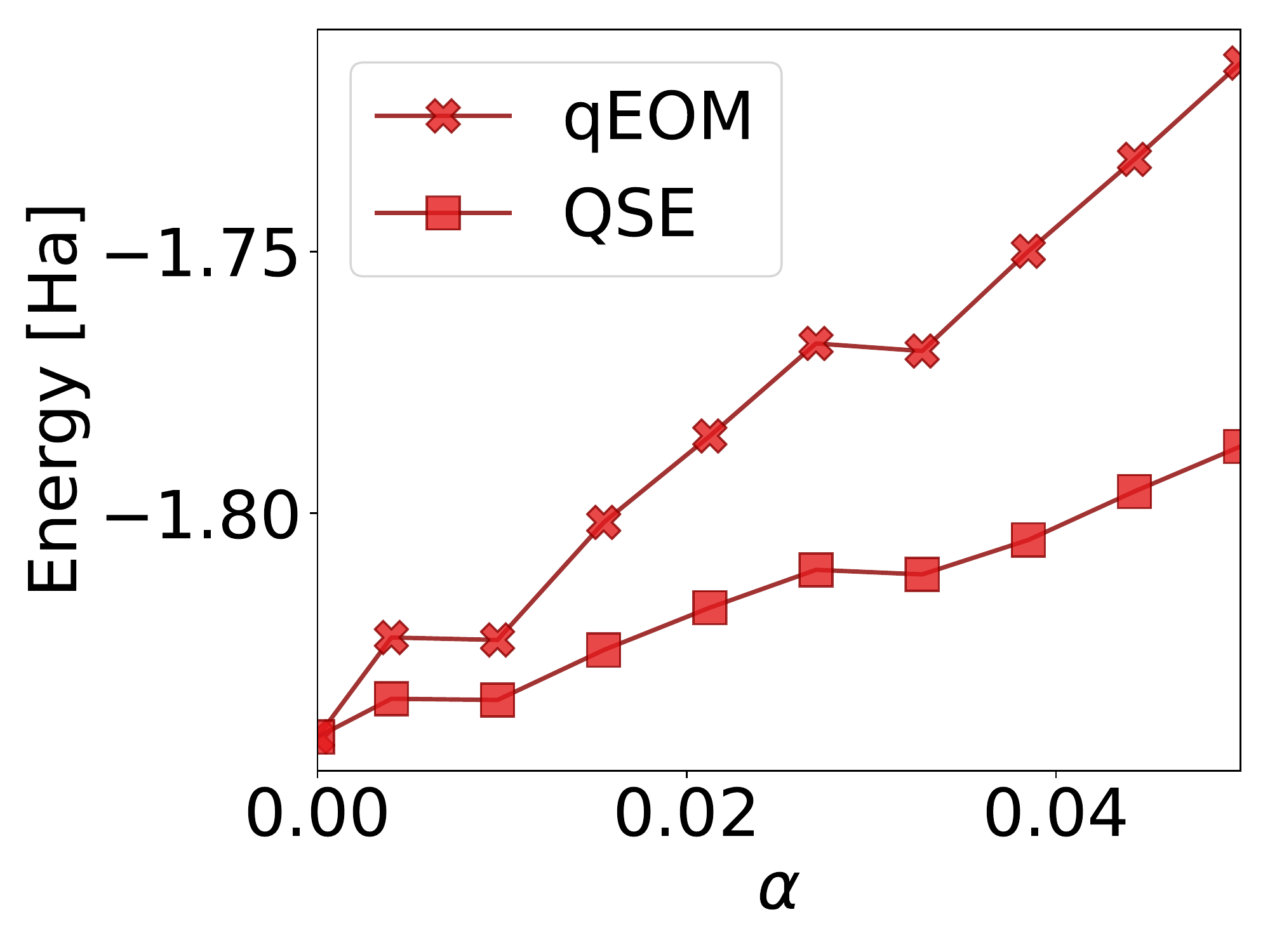}}
  \hspace{1cm}
  \subfloat{\includegraphics[width = 0.4\textwidth]{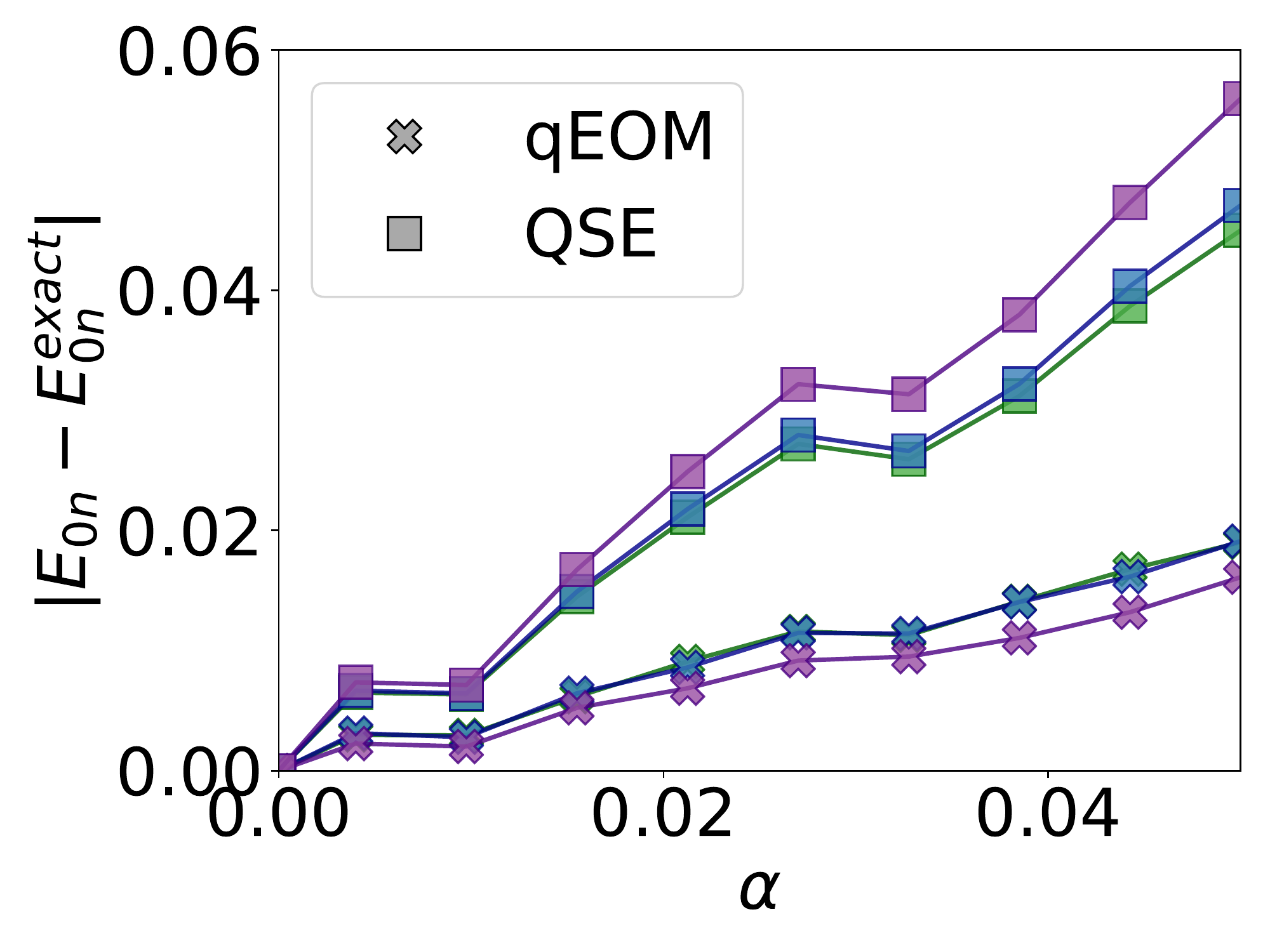}}
    \caption{Left: Ground state energy of H$_2$ (4 qubits) at equilibrium bond length as a function of the $\alpha$ parameter. In the case of qEOM the energy is the expectation value of the Hamiltonian with $\ket{\Psi} = \text{U}(\alpha)\ket{\Psi_0}$. In the other case, it is the minimum energy found after solving the QSE equations. Each point is the average over 10'000 trials. Right: The error in excitation energies for the first (green), second (blue) and third (purple) excitation energies of H$_2$ (4 qubits) at equilibrium bond length. Each point is the average over 10'000 trials.}
    \label{eomvsqse}
\end{figure}
Differently from the method introduced in this work, the QSE approach discussed in~\cite{Mcclean2017} neglects the de-excitation operators (as in the Tamm-Dancoff approximation~\cite{Fetter2012, Ring2004}), which limits its application to systems for which this approximation is valid (e.g., far from conical intersections, see for instance~\cite{shu2017}). 
Moreover, QSE includes the identity operator in the pool of `excitation' operators. Consequently, QSE is not size-intensive, thus not generally applicable in the calculation of energy differences.
Additionally, from the observations of the previous section, we conclude that the accuracy in the calculation of the ground state is crucial for recovering excitation energies within chemical accuracy. 
With the current statistical and hardware noise levels, obtaining such precision is still very challenging and, therefore, we require robust excited states algorithms that can cope with these experimental conditions.
We assess the performance of qEOM and QSE by computing the excitation energy gaps (which are in general of interest rather than absolute energies) for the H$_2$ molecule (STO-3G basis set, 4 qubits). 
We consider the following ground state:
\begin{equation}
\ket{\Psi} = \text{U}(\alpha)\ket{\Psi_0}
\end{equation}
where $\ket{\Psi_0}$ is the exact ground state obtained by diagonalization of the Hamiltonian, and U is a random unitary matrix characterized by a density $\alpha$ of off-diagonal elements. The results are averaged over 10'000 trials. The left-hand side of Fig.~\ref{eomvsqse} shows the ground state energy. For $\alpha = 0$, the energy is exact (since $U(0)$ is the identity matrix). For the qEOM the ground state energy corresponds to the expectation value of the Hamiltonian: $\bra{\Psi}\text{H}\ket{\Psi}$. In the case of QSE, the ground state energy is obtained from the solution of the pseudo-eigenvalue problem as for the excitation energies. 
Note that in this work for both methods (qEOM and QSE) all (spin number conserving) single- and double-excitation operators are considered (for QSE the identity operator is also included). 
In our calculations of the excitation energies of $H_2$, qEOM is more robust than QSE against the different source of noise (\textit{e.g.} gate noise or decoherence effects) in the preparation of the ground state wave function.
\section{Molecular Hamiltonians and circuits}
\subsection{Hydrogen}
\begin{table}[h]
   \centering
   \begin{tabular}{|m{3.5cm}|m{1.6cm}|m{1.6cm}|m{1.6cm}|m{1.6cm}|}
   \hline
        \makecell{IIIZ\\0.169885\\IIZI\\-0.21886\\IZII\\0.169885\\ZIII\\-0.21886\\IZIZ\\0.168212}\makecell{IIZZ\\0.120051\\IZZI\\0.165494\\ZIIZ\\0.165494\\ZIZI\\0.173954\\ZZII\\0.120051} & \makecell{XXYY\\0.045443} & \makecell{YYXX\\0.045443} & \makecell{XXXX\\0.045443} & \makecell{IIII\\1.006421\\YYYY\\0.045443} \\
    \hline
   \end{tabular}
   \caption{The H$_{2}$ Hamiltonian at equilibrium bond length (0.75\AA).  Listed are all the Pauli operators with the corresponding coefficients, not taking into account for the energy shifts due to the frozen core orbitals and the Coulomb repulsion between nuclei. Each column corresponds to a different tensor product basis set. $X$, $Y$, $Z$, $I$  here  stand  for  the  Pauli  matrices $\sigma_x$, $\sigma_y$, $\sigma_z$ and  the  identity operator on a single qubit subspace, respectively.}
\end{table}
\subsection{Lithium Hydride}
\label{section:appendix_reduction}
\begin{table}[ht]
   \centering
   \begin{tabular}{|m{1.6cm}|m{1.6cm}|m{1.6cm}|m{1.6cm}|m{1.6cm}|m{1.6cm}|m{1.6cm}|m{1.6cm}|m{1.6cm}|}
   \hline
        \makecell{IIII \\ 0.681524 \\IIYY \\ 0.038896 \\ YYII \\ 0.038896 \\ YYYY \\ 0.030982} & \makecell{XXII\\0.038896\\XXYY\\0.030982} & \makecell{IIXX\\0.038896\\YYXX\\0.030982} & \makecell{XXXX\\0.030982} & \makecell{IZII\\0.428555\\ZIII\\0.166096\\IZYY\\-0.03169\\ZIYY\\-0.03355\\ZZII\\0.082479} & \makecell{IZXX\\-0.03169\\ZIXX\\-0.03355} & \makecell{IIIZ\\0.428555\\IIZI\\0.166096\\YYIZ\\-0.03169\\IIZZ\\0.082479\\YYZI\\-0.03355} & \makecell{XXIZ\\-0.03169\\XXZI\\-0.03355} & \makecell{IZZI\\0.113461\\ZIIZ\\0.113461\\ZIZI\\0.113447\\IZIZ\\0.121828}\\
    \hline
   \end{tabular}
   \caption{The LiH Hamiltonian at equilibrium bond length (1.6\AA) after reduction of the active space. Listed are all the Pauli operators with the corresponding coefficients, not taking into account for the energy shifts due to the frozen core orbitals and the Coulomb repulsion between nuclei. Each column corresponds to a different tensor product basis set. $X$, $Y$, $Z$, $I$  here  stand  for  the  Pauli  matrices $\sigma_x$, $\sigma_y$, $\sigma_z$ and  the  identity operator on a single qubit subspace, respectively}
\end{table}
The UCCSD circuit of LiH requires over 12000 CNOT gates and the optimization of 92 parameters.
The circuit depth can, however, be reduced by following two steps. Firstly, the UCC ansatz is restricted to one excitation i.e., one variational parameter. This excitation is selected by computing the MP2 coefficients of the CI expansion, 
\begin{equation}
    C_{ijlk}^{\text{MP2}}=\frac{h_{ijkl}-h_{ijlk}}{e_i+e_j-e_k-e_l},
\end{equation}
where, $h_{ijkl}$ are the two-electron integrals and $e_i$ is the energy of orbital $i$. We select the excitation with the largest MP2 coefficient. 
By including only the relevant orbitals, we also reduce the 10-qubit LiH active space to a 4-qubit space. Indeed, we consider the active space to be represented by a 4-qubit quantum register while the inert space is mapped to a 6-qubit register, the active and inert registers are uncorrelated and therefore,
\begin{equation}
    \bra{\psi_{\text{full}}}\hat{\text{A}}\hat{\text{I}}\ket{\psi_{\text{full}}}=\bra{\psi_{\text{active}}}\hat{\text{A}}\ket{\psi_{\text{active}}}\bra{\psi_{\text{inert}}}\hat{\text{I}}\ket{\psi_{\text{inert}}}
\end{equation}
where, $\hat{\text{A}}$ and $\hat{\text{I}}$ are operators acting on the active and inert space respectively. In the inert register, since the qubits are all in the $\ket{0}$ state, $\langle Z \rangle = \langle I \rangle = 1$ and $\langle X \rangle = \langle Y \rangle = 0$. It follows that only the active register has to be modeled in quantum hardware. Let us illustrate the previous paragraph with a short example. In the 10-qubit LiH, the four first qubits represent the active space while the six last ones are inert. Let's consider IZIZIIXIII to be one of the PS composing the Hamiltonian. In this example, IZIZ is measured on the active register while IIXIII is evaluated on the inert register. Since one of the qubits of the inert register is measured in the X basis, $\langle \text{IIXIII} \rangle_{\text{inert}} = 0$ and therefore, $\langle \text{IZIZIIXIII} \rangle_{\text{full}} = 0$. This PS can be set to 0. On the other hand, if we consider IZIZIIZIII, $\langle \text{IIZIII} \rangle_{\text{inert}} = 1$ and the value of the PS can be set to that measured on the active register: $\langle \text{IZIZIIZIII} \rangle_{\text{full}} = \langle \text{IZIZ} \rangle_{\text{active}}$. Thus, by reducing the active space we can measure the 10-qubit LiH Hamiltonian on a 4-qubit register.
\begin{figure}[t]
\includegraphics[width = 0.4\columnwidth]{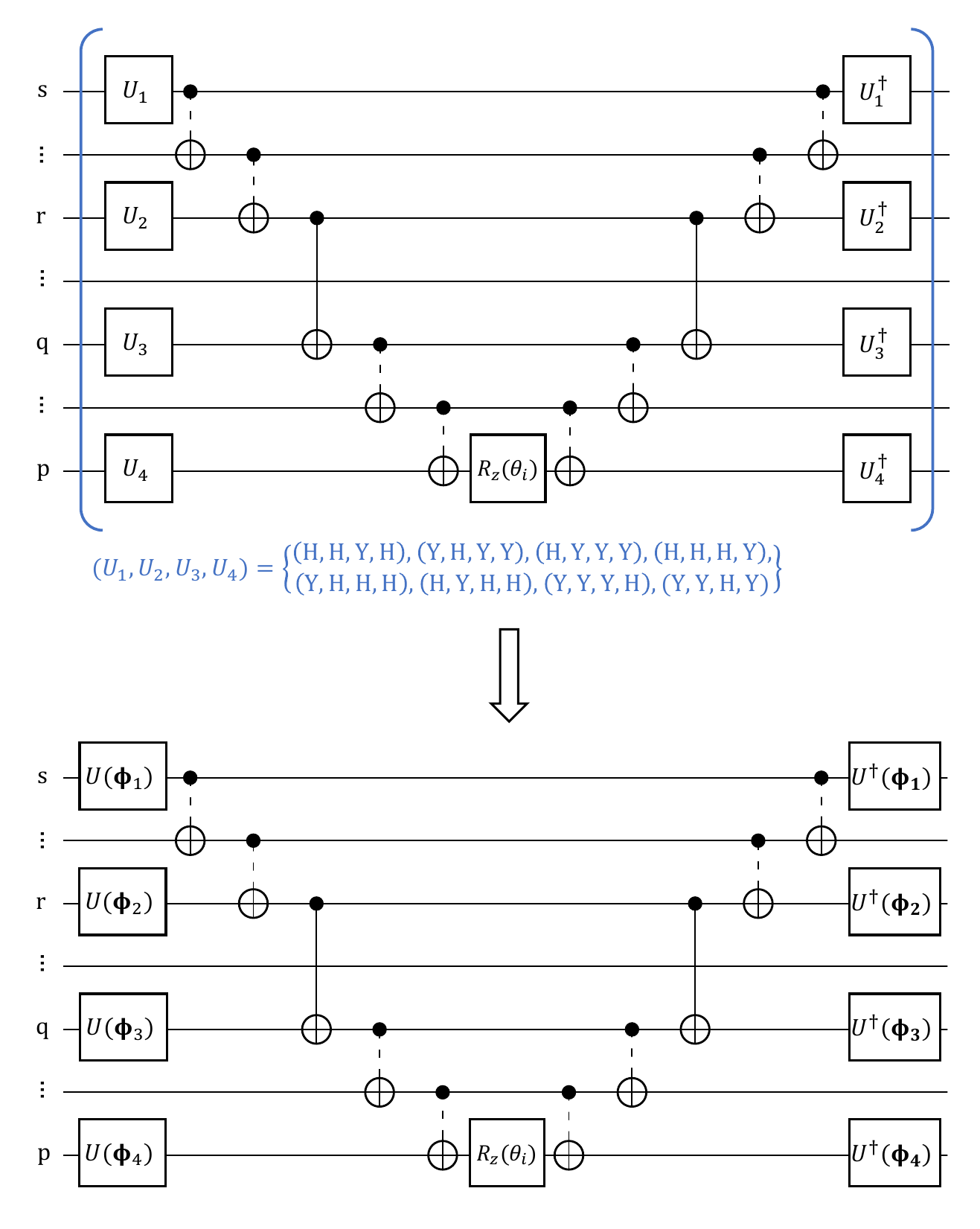}
\caption {Reduction of the usual eight blocks of the double excitation part of the UCCSD circuit to a single block.}
\label{fig:UCC2hUCC}
\end{figure}

\begin{figure}[ht]
\includegraphics[width = 0.5\columnwidth]{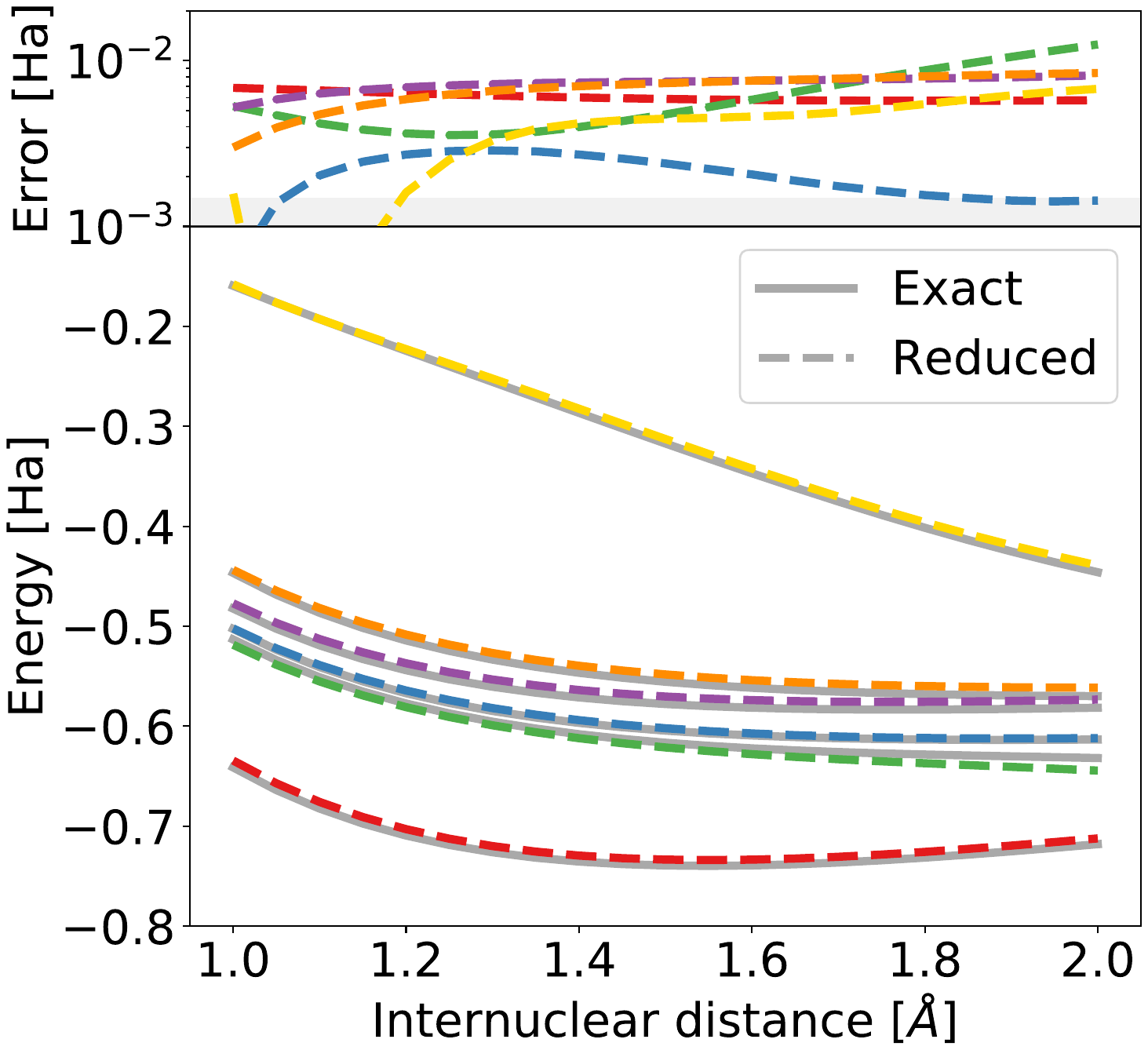}
\caption {Lower panel: Dissociation profile the $\rm{LiH}$ molecule. The grey lines represent the exact eigenstates of the Hamiltonian obtained from its diagonalizaton. The coloured-dashed lines show the six lowest-lying electronic states obtained with VQE and qEOM in statevector simulation with the reduced circuit. 
Upper panel: Corresponding energy errors along the dissociation profile. The gray shaded area corresponds to the energy range within chemical accuracy}
\label{fig:lih_sim}
\end{figure}

The second step to reduce the circuit depth consists of modifying the UCC circuit. In the regular UCC, the circuit (derived by applying the Jordan-Wigner transformation on the classical UCC Ansatz) representing a single excitation is made up of four entangling blocks, with different pre/post rotations. Those four blocks are parametrized with the same angle $\theta$. For the double excitations, a similar construction is used and extended to eight entangling blocks. All pre/post rotations are fixed and different for each entangling block. We propose to replace the four and eight entangling blocks by a single block in which the pre/post rotations angles $\{\bm{\Phi}\}$ are optimized on a small system (e.g., H$_2$ is used to optimize the pre/post rotation angles of the double excitation blocks), see Fig.~\ref{fig:UCC2hUCC}. 
For the 4-qubit LiH circuit, this allows us to further reduce the circuit from 48 to 6 CNOTs without loss of accuracy.
This circuit can recover from 57\% of the correlation energy at 1.0 \AA\ up to 80\% at 2.0 \AA\ (the amount of correlation energy captured by this circuit along the dissociation curve, is given in Table~\ref{fig:E_CORR_REC}). We assume this approximation of the ground state to be good enough to compute accurate excitation energies.\\
\begin{table}[ht]
    \centering
    \begin{tabular}{|c|c|c|c|}
    \hline
    Bond Length & $E_{\text{corr}}$ VQE [Ha] & $E_{\text{corr}}$ Exact [Ha] & $E_{\text{corr}}$ recovered (\%) \\
    \hline
         1.0	&	-0.009797	&	-0.017098	&	57.3	\\
1.1	&	-0.009841	&	-0.016794	&	58.6	\\
1.2	&	-0.010169	&	-0.016815	&	60.5	\\
1.3	&	-0.010782	&	-0.017186	&	62.7	\\
1.4	&	-0.011688	&	-0.017915	&	65.2	\\
1.5	&	-0.012899	&	-0.019005	&	67.9	\\
1.6	&	-0.014430	&	-0.020460	&	70.5	\\
1.7	&	-0.016299	&	-0.022289	&	73.1	\\
1.8	&	-0.018525	&	-0.024505	&	75.6	\\
1.9	&	-0.021132	&	-0.027129	&	77.9	\\
2.0	&	-0.024142	&	-0.030182	&	80.0	\\
2.2	&	-0.031456	&	-0.037689	&	83.4	\\
2.5	&	-0.045769	&	-0.052850	&	86.6	\\
\hline
    \end{tabular}
    \caption {Percentage of correlation energy recovered by the reduced 4-qubit active space UCC-inspired circuit for LiH through the dissociation profile. The VQE results are obtained with statevector-type simulations, the reduced circuit and the COBYLA optimizer. The exact results are obtained by diagonalization of the full HF/STO-3G Hamiltonian.}
\label{fig:E_CORR_REC}
    \label{tab:my_label}
\end{table}
Regardless of the method chosen to approximate the ground state, we want to compute the excited states of the LiH molecule in the STO-3G (10-qubit) basis. The ground state wave function is evolved in the reduced active space, and the previously described measurement method is applied to compute the qEOM operators and reconstruct the full pseudo-eigenvalue problem. The expected accuracy, given the reduced circuit, is depicted in Fig.~\ref{fig:lih_sim} and lies between 1 and 10 mHa.
\section{Experimental details}
\label{section:exp_details}
\begin{figure}[ht]
\includegraphics[width = 0.9\columnwidth]{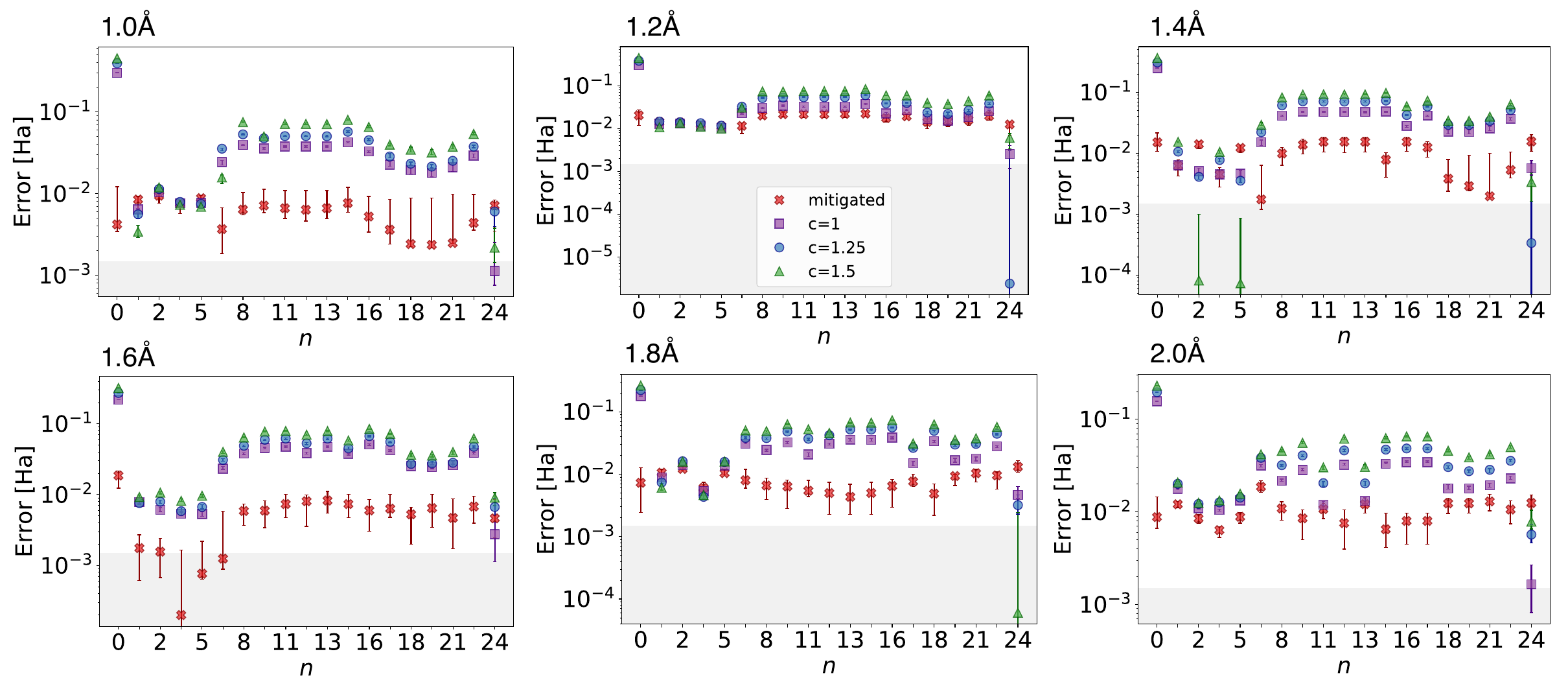}
\caption {Error in the experimentally obtained ground state energies ($n=0$) and excitation energies up to $n=24$, for several internuclear distances. The error is computed with respect to the results obtained by statevector-type simulation of the reduced circuit. The results are shown for each of the three stretch factors and after mitigation. The energy of the gaps grows from left to right. The gray shaded area corresponds to the energy range within chemical accuracy.  The error bars are computed using 50 numerical experiments obtained by bootstrapping of the experimental data points, and depict the range between the 1st and 3rd quantile. For all the bond lengths, the excitation energies are seen to be more robust than the ground state energies, at various noise levels (i.e. stretch factors).}
\label{fig:lil_all_states}
\end{figure}
\end{document}